\documentclass{moriond}

\usepackage{amssymb}                             
\usepackage{amsmath}        
\usepackage{grffile} 
\usepackage{slashed}

\def\simge{
    \mathrel{\rlap{\raise 0.511ex
        \hbox{$>$}}{\lower 0.511ex \hbox{$\sim$}}}}
\def\simle{
    \mathrel{\rlap{\raise 0.511ex 
        \hbox{$<$}}{\lower 0.511ex \hbox{$\sim$}}}}
        
\def\be{\begin{equation}}
\def\ee{\end{equation}}
\def\bea{\begin{eqnarray}}
\def\eea{\end{eqnarray}}

\newcommand{\Photo}{\includegraphics[height=35mm]{mypicture}}

\begin{document}
\vspace*{4cm}
\title{Hadronic light-by-light contribution to the muon anomalous magnetic moment from lattice QCD}


\author{Thomas Blum and Luchang Jin}
\address{Physics Department, University of Connecticut, 2152 Hillside Road, Storrs, CT, 06269-3046, USA\\RIKEN BNL Research Center, Brookhaven National Laboratory, Upton, New York 11973, USA}
\author{Norman Christ}
\address{Physics Department, Columbia University, New York, New York 10027, USA}
\author{Masashi Hayakawa}
\address{Department of Physics, Nagoya University, Nagoya 464-8602, Japan\\Nishina Center, RIKEN, Wako, Saitama 351-0198, Japan}
\author{Taku Izubuchi}
\address{Physics Department, Brookhaven National Laboratory, Upton, New York 11973, USA\\RIKEN BNL Research Center, Brookhaven National Laboratory, Upton, New York 11973, USA}
\author{Chulwoo Jung}
\address{Physics Department, Brookhaven National Laboratory, Upton, New York 11973, USA}
\author{Christoph Lehner}
\address{Universit\"at Regensburg,
Fakult\"at f\"ur Physik, 93040, Regensburg, Germany\\Physics Department, Brookhaven National Laboratory, Upton, New York 11973, USA}

\renewcommand{\Photo}{}

\maketitle\abstracts{We report preliminary results for the hadronic light-by-light scattering contribution to the muon anomalous magnetic moment. Several ensembles using 2+1 flavors of M\"obius domain-wall fermions, generated by the RBC/UKQCD collaborations, are employed to take the continuum and infinite volume limits of finite volume lattice QED+QCD. We find $a_\mu^{\rm HLbL} = (7.41\pm6.33)\times 10^{-10}$.
}

\section{Introduction}

The anomalous magnetic moment of the muon is providing an important test of the Standard Model. An ongoing experiment at Fermilab (E989) and one planned at J-PARC (E34) aim to reduce the experimental uncertainty by a factor of four, and \href{http://www.int.washington.edu/PROGRAMS/19-74W/}{similar efforts} are underway on the theory side. A key part of the latter is to compute the hadronic light-by-light (HLbL) contribution from first principles using lattice QCD. Such a calculation, with all errors under control, leaves no room for doubt when the ultimate comparison arrives.

\begin{figure}
    \centering
    \includegraphics[width=0.3\textwidth]{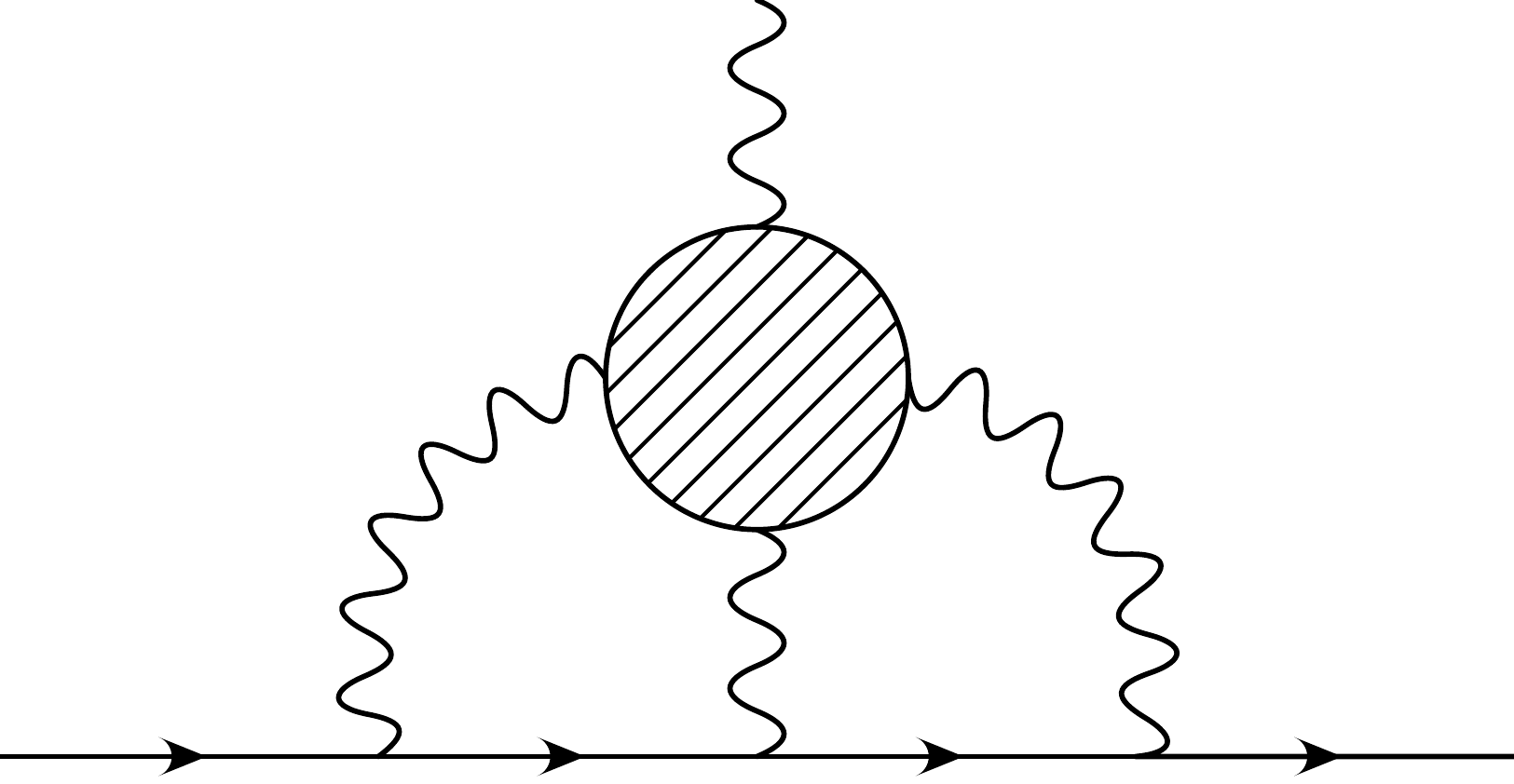}~~~+~~~\includegraphics[width=0.3\textwidth]{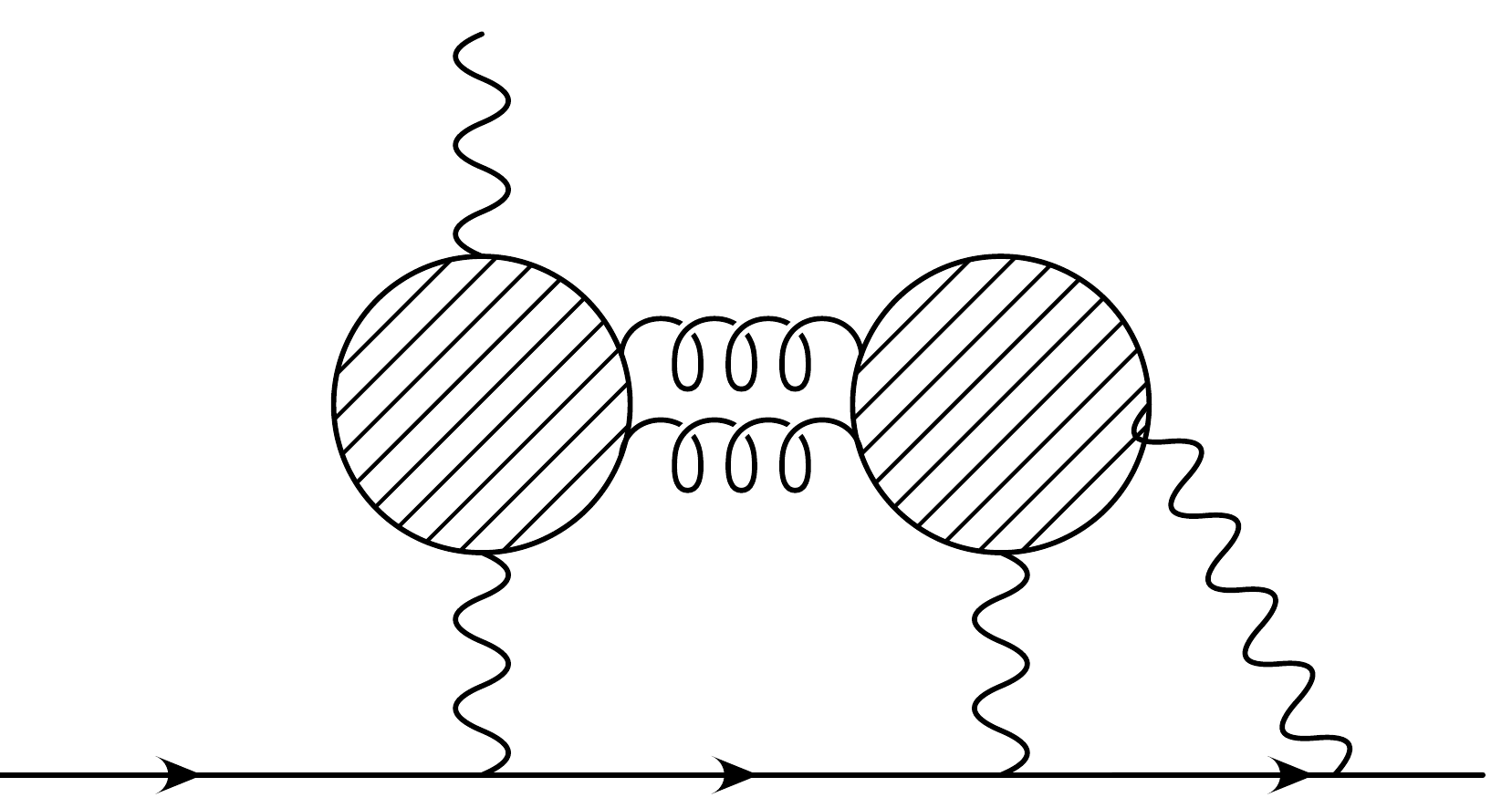}~~~$+~~~\cdots$
    \caption{Leading contributions from hadronic light-by-light scattering to the muon anomaly. The shaded circles represent quark loops containing QCD interactions to all orders. Horizontal lines represent muons. Quark- connected (left) and disconnected (right) diagrams are shown. Ellipsis denote diagrams obtained by permuting the photon contractions with the muons and diagrams with three and four quark loops with photon couplings.}
    \label{fig:hlbl feyn diags}
\end{figure}

The anomalous magnetic moment is an intrinsic property of a spin-1/2 particle, and is defined through its interaction with an external magnetic field. Lorentz and gauge symmetries dictate the form of the interaction,
\begin{eqnarray}
\langle \mu (\vec p^\prime) | J_\nu(0) |\mu(\vec p)\rangle &=&
-e \bar u(\vec p^\prime)\left(F_1(q^2)\gamma_\nu+i\frac{F_2(q^2)}{4 m}[\gamma_\nu,\gamma_\rho] q_\rho\right)u(\vec p),
\label{eq:ff}
\end{eqnarray}
where $J_\mu$ is the electromagnetic current, and $F_1$ and $F_2$ are form factors, giving the charge and magnetic moment at zero momentum transfer ($q=p^\prime-p=0$). The anomalous part of the magnetic moment is given by $F_2(0)$ alone,
\begin{eqnarray}
a_\mu &\equiv& (g-2)/2 = F_2(0).
\end{eqnarray}
The desired matrix element in (\ref{eq:ff}) is conventionally extracted in quantum field theory from a correlation function of fields as depicted in Fig.~\ref{fig:hlbl feyn diags}. Here we work in coordinate (Euclidean) space and use Lattice QCD for the hadronic part which is intrinsically non-perturbative. QED is treated in two ways, first on a discrete, finite, lattice (QED$_L$) and second in the continuum and infinite volume (QED$_\infty$). Note that we always work in a perturbative framework with respect to QED, $i.e$, only diagrams where the hadronic part is connected to the muon by three photons enter the calculation.

\section{QED$_L$ Method}

\begin{figure}
    \centering
    \includegraphics[width=\textwidth]{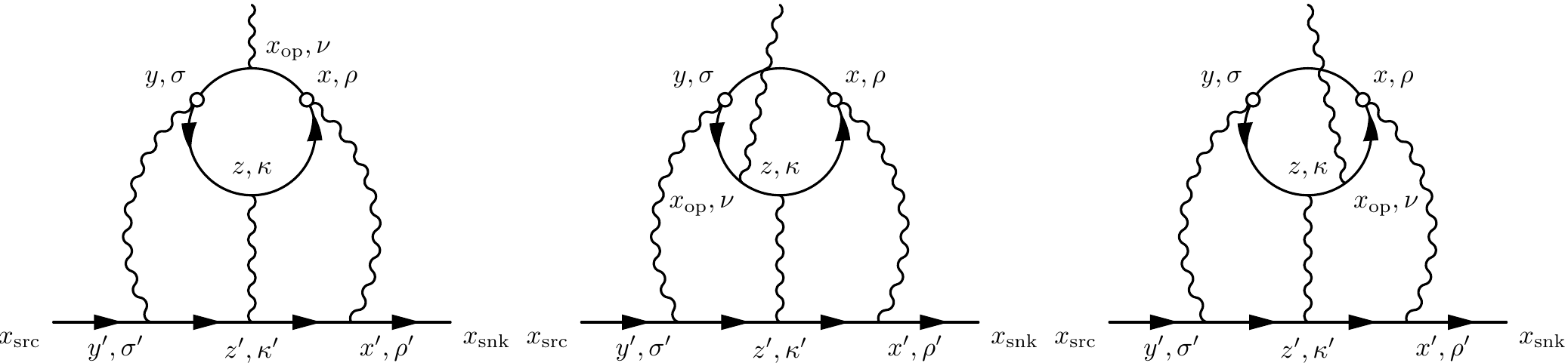}
    \caption{Correlation functions. Sums over $x$ and $y$ are computed stochastically. The third internal vertex $z$ and the external vertex $x_{\rm op}$ are summed over exactly. The sums on the muon line are done exactly using FFT's. Strong interactions to all orders are not shown.}
    \label{fig:point src method}
\end{figure}

Here the muon, photons, quarks, and gluons are treated on a single finite, discrete lattice. The method is described in great detail in Ref.~\cite{Blum:2015gfa}, and the (quark-connected) diagrams to be computed are shown in Fig.~\ref{fig:point src method}. It is still not possible to do all of the sums over coordinate space vertices exactly with currently available compute resources. Therefore we resort to a hybrid method where two of the vertices on the hadronic loop are summed stochastically: point source propagators from coordinates $x$ and $y$ are computed, and their sink points are contracted at the third internal vertex $z$ and the external vertex $x_{\rm op}$. Since the propagators are calculated to all sink points, $z$ and $x_{\rm op}$ can be summed over the entire volume. The sums over vertices $x$ and $y$ are then done stochastically by computing many ($O(1000)$) random pairs of point source propagators. To do the sampling efficiently, the pairs are chosen with an empirical distribution designed to frequently probe the summand where it is large, less frequently where it is small. Since QCD has a mass-gap, we know the hadronic loop is exponentially suppressed according to the distance between any of the vertices, including $|x-y|$. As we will see, the main contribution comes from distances less than about 1 fm. The muon line and photons are computed efficiently using FFT's; however, because they must be calculated many times, the cost is not trivial.

Two additional, but related, parts of the method bear mentioning. First, the form dictated by the right hand side of Eq.~\ref{eq:ff} suggests the limit $q\to0$ is unhelpful since the desired $F_2$ term is multiplied by 0. Second, in our Monte Carlo lattice QCD calculation the error on the $F_2$ contribution blows up in this limit. The former is avoided by evaluating the first moment with respect to $\vec{x}_{\mathrm{op}}$ at the external vertex and noticing that an induced extra term vanishes exponentially in the infinite volume limit~\cite{Blum:2015gfa}. This moment method allows the direct calculation of the correlation function at $q=0$, and hence $F_2(0)$. The second issue is avoided by enforcing the Ward Identity exactly on a configuration-by-configuration basis, $i.e.$, before averaging over gauge fields. This makes the factor of $q$ in Eq.~(\ref{eq:ff}) exact for each measurement and not just in the average. The Ward Identity is enforced by inserting the external photon at all possible locations on the quark loop. The three distinct possibilities are shown in Fig.~\ref{fig:point src method}. By the way, it is the Ward Identity that guarantees the unwanted term in the moment method vanishes.
Implementing the above techniques produces an order $O(1000)$ fold improvement in the statistical error over the  original non-perturbative method for the hadronic light-by-light scattering contribution~\cite{Blum:2014oka}.

\subsection{disconnected diagrams}

\begin{figure}
    \centering
\includegraphics[width=0.25\textwidth]{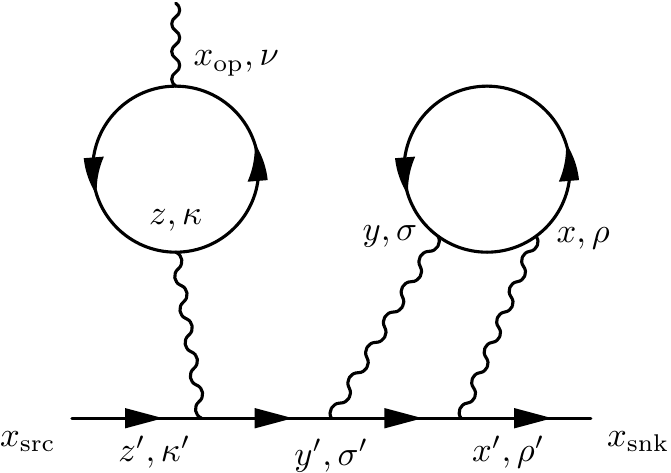}
\includegraphics[width=0.25\textwidth]{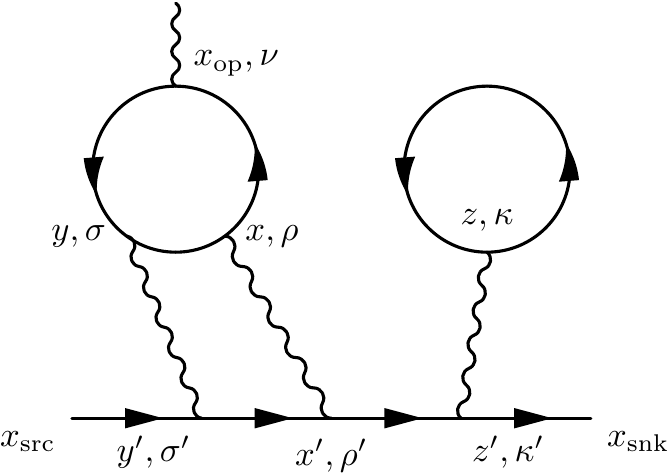}
\includegraphics[width=0.25\textwidth]{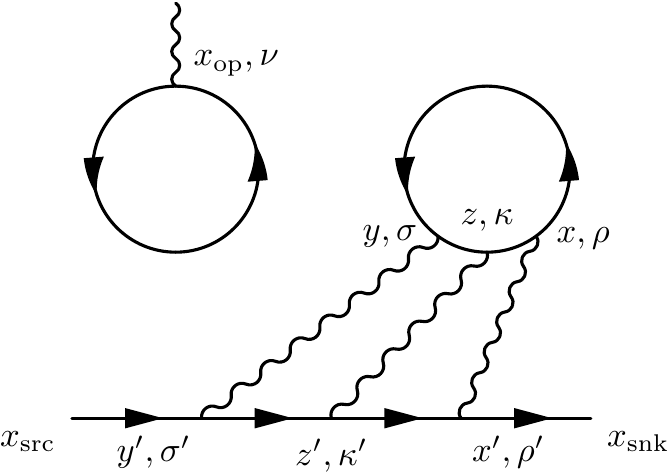}\\
\includegraphics[width=0.25\textwidth]{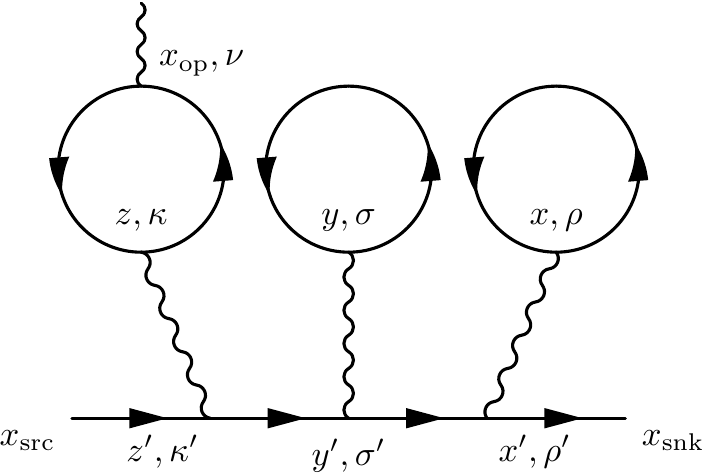}
\includegraphics[width=0.25\textwidth]{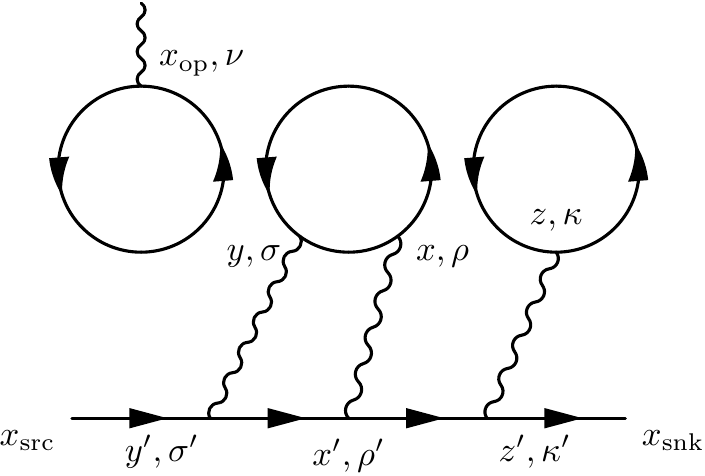}
\includegraphics[width=0.275\textwidth]{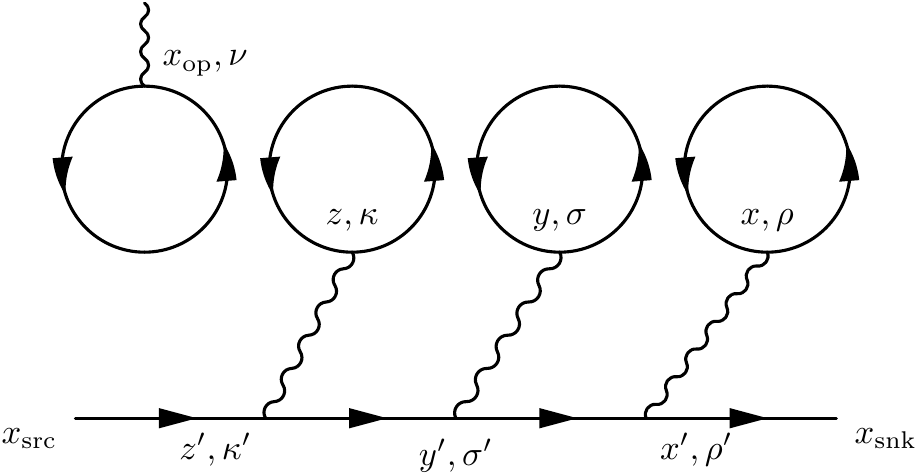}
    \caption{Disconnected diagrams contributing to the muon anomaly. The top leftmost is the leading one, and does not vanish in the $SU(3)$ flavor limit. Strong interactions to all orders, including gluons connecting the quark loops, are not shown.}
    \label{fig:disco diags}
\end{figure}

The quark-disconnected diagrams that occur at $O(\alpha^3)$ are shown Fig.~\ref{fig:disco diags}). All but the upper-leftmost diagram vanish in the $SU(3)$ flavor limit and are suppressed by powers of $m_{u,d} - m_s$ depending on the number of loops with a single photon attached. For now we ignore them and concentrate on the leading diagram which is computed with a method similar to the one described in the previous section~\cite{Blum:2016lnc}.

To ensure loops are connected by gluons, explicit vacuum subtraction is required. However, in the leading diagram the moment at $x_{\rm op}$ implies that the left-hand loop in Fig.~\ref{fig:disco diags} vanishes due to parity symmetry, and the vacuum subtraction is done to reduce noise. 

As for the connected case, two point sources (at $y$ and $z$ in Fig.~\ref{fig:disco diags}) are chosen randomly, and the sink points sinks are summed over. $M$ point source propagators are computed, and all $M^2$ combinations are used to perform the stochastic sum. This ``$M^2$ trick" is crucial to bring the statistical fluctuations of the disconnected diagram under control. 

\subsection{lattice setup}

The simulation parameters are given in Tab.~\ref{tab:ensembles}. All particles have their physical masses (but not including isospin breaking for the up and down quark masses). The discrete Dirac operator is known as the (M\"obius) domain wall fermion ((M)DWF)) operator. Similarly the discrete gluon action is given by the plaquette plus rectangle Iwasaki gauge action. Three ensembles with larger lattice spacing employ the dislocation-suppressing-determinant-ratio (DSDR) to soften explicit chiral symmetry breaking effects for MDWF. 

The muons and photons take discrete free-field forms. The muons are DWF with infinite size in the extra fifth dimension, and the photons are non-compact in the Feynman gauge. In the latter all modes with $\vec q=0$ are dropped, a finite volume formulation of QED known as QED$_L$~\cite{Hayakawa:2008an}.

\begin{table}[htp]
\begin{center}
\begin{tabular}{|c|c|c|c|c|c|}
\hline
& 48I & 64I & 24D & 32D & 48D \\
\hline
$a^{-1}$ (GeV) & 1.73 & 2.359 & 1.015 & 1.015 & 1.015\\
$a$ (fm) & 0.114&0.084 & 0.2 & 0.2  & 0.2 \\
$L$ (fm)  & 5.47 & 5.38 & 4.8 & 6.4 & 9.6\\
$L_s$ & 48 & 64 & 24 & 24  & 24\\
$m_\pi$ (MeV) & 139 & 135 & 140 & 140  & 140\\
$m_\mu$ (MeV) & 106 & 106 & 106 & 106  & 106\\
\# meas (conn., disc.) & 65, 99 & 43, 44 & 87, 80 & 64, 68 & 62, 0\\
\hline
\end{tabular}
\caption{2+1 flavors of MDWF gauge field ensembles generated by the RBC/UKQCD collaborations~\protect\cite{Blum:2014tka}. 
}
\end{center}
\label{tab:ensembles}
\end{table}

\subsection{test in pure QED}
 
\begin{figure}
     \centering
     \includegraphics[width=0.5\textwidth]{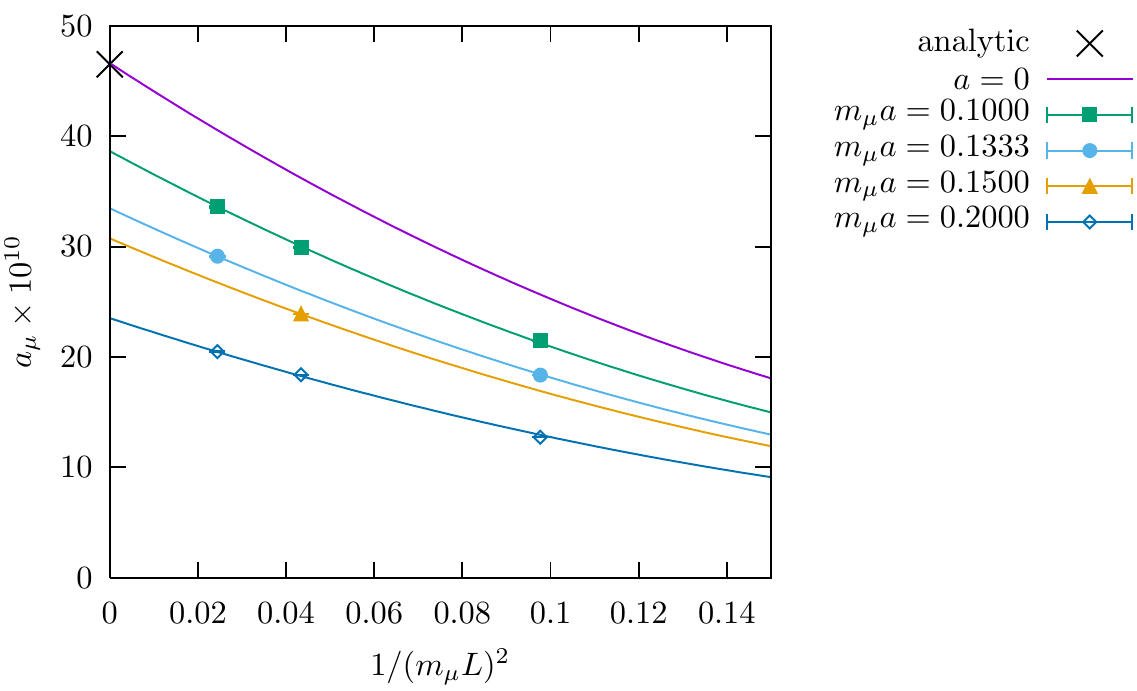}
     \caption{QED light-by-light scattering contribution to the muon anomaly. Lattice spacing decreases from bottom to top. Solid lines are from a fit using Eq.~(\ref{eq:qed extrap}).}
     \label{fig:qed test}
 \end{figure}

 Before moving to the hadronic case, we tested the method in pure QED~\cite{Blum:2015gfa}. Results for several lattice spacings and box sizes are shown in Fig.~\ref{fig:qed test}. The systematic uncertainties are large, but under control. Note that the finite volume errors are polynomial in $1/L$ and not exponential. The data are well fit to the form
\begin{equation}
F_2(a,L)=F_2\left(1-\frac{b_1}{(m_\mu L)^2}+\frac{b_2}{(m_\mu L)^4}\right)(1-c_1 a^2+c_2 a^4).
\label{eq:qed extrap}
\end{equation}
The continuum and infinite volume limit is $F_2(0)=46.6(2) \times 10^{-10}$ for the case where the lepton mass in the loop is the same as the muon mass, which is quite consisent with the well known perturbative value~\cite{Laporta:1991zw}, $46.5\times10^{-10}$.

\subsection{results for QCD}

Our physical point calculation~\cite{Blum:2016lnc} started on the $48^3$, $a^{-1}=1.730$ GeV, Iwasaki ensemble listed in the first column of Tab.~\ref{tab:ensembles}, for which we found 
$a_\mu^{\rm cHLbL} = (11.60\pm0.96)\times 10^{-10}$,
$a_\mu^{\rm dHLbL} = (-6.25\pm0.80)\times 10^{-10}$,
 and $a_\mu^{\rm HLbL} = (5.35\pm1.35)\times 10^{-10}$ for the connected, leading disconnected, and total HLbL contributions to the muon anomaly, respectively. The errors quoted are purely statistical.
We have since improved the statistics on the leading disconnected diagram with measurements on 34 additional configurations, and the contribution becomes $-6.15 (61)\times 10^{-10}$. Since then we have computed on several additional ensembles in order to take the continuum and infinite volume limits (see Tab.~\ref{tab:ensembles}). We next computed on a $64^3$, $a^{-1}=2.359$ GeV, companion to the original $48^3$ Iwasaki ensemble with roughly the same volume. This allows a continuum limit at finite volume, $a_\mu^{\rm cHLbL} =16.94(3.78)$, $a_\mu^{\rm dHLbL}=-12.29(3.35)$, and $a_\mu^{\rm HLbL}=4.66(4.39)$. Notice there is a large cancellation between the connected and disconnected diagrams that persists for $a\to0$. Even though the individual contributions are relatively well resolved, the total is not. The cancellation is expected since hadronic light-by-light scattering in this case is dominated by the $\pi^0$ which contributes to both diagrams, but with opposite sign~\cite{Bijnens:2016hgx,Jin:2016rmu,Gerardin:2017ryf}. Notice also that the $a^2$ corrections are individually large
 but also tend to cancel in the sum. 
 
\begin{figure}
    \centering
    \includegraphics[width=0.3\textwidth]{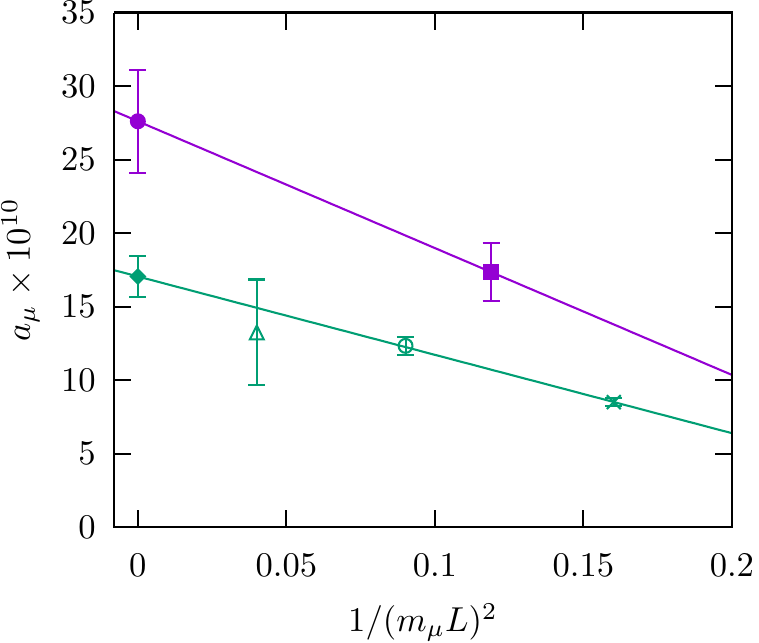}
    \includegraphics[width=0.3\textwidth]{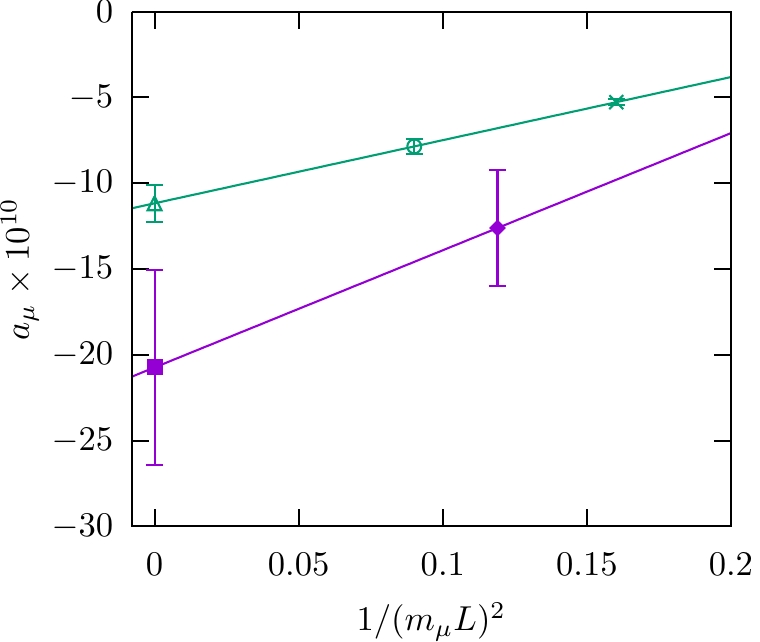}
    \includegraphics[width=0.3\textwidth]{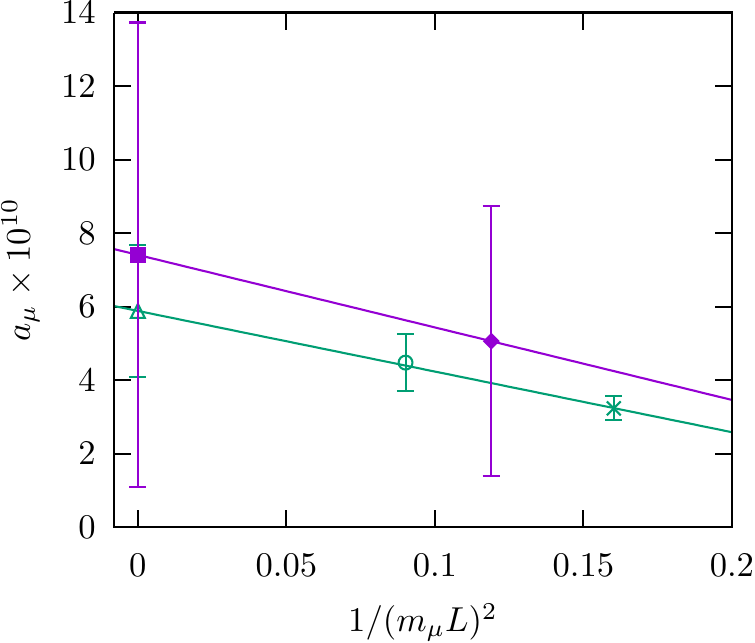}
    \caption{Infinite volume extrapolation. Connected (left), disconnected (middle) and total (right).}
    \label{fig:qedl extrap}
\end{figure}

 Next the infinite volume limit must be taken. To do this we add another set of ensembles with a slightly different gauge action (Iwasaki+DSDR) and larger lattice spacing so that large physical volumes can be realized (roughly 4.8, 6.4, and 9.6 fm boxes). See Tab.~\ref{tab:ensembles} for details. The results are displayed in Fig.~\ref{fig:qedl extrap} along with curves obtained from Eq.~(\ref{eq:qed extrap}) with $b_2=c_2=0$.
We first extrapolate the two Iwasaki ensembles to $a\to 0$, as before, then combine with the I-DSDR ensembles to take the infinite volume limit. We find for the connected, disconnected, and total contributions,
$a_\mu^{\rm cHLbL} = (27.16\pm6.25)\times 10^{-10}$,
$a_\mu^{\rm dHLbL} = (-20.20\pm5.65)\times 10^{-10}$,
$a_\mu^{\rm HLbL} = (6.96\pm7.40)\times 10^{-10}$, respectively.
Similar to the non-zero lattice spacing errors, there are large finite volume corrections for the individual contributions, which again largely cancel in the sum.

While the large relative error on the total is a bit unsatisfactory, we emphasize that our result represents an important estimate on the hadronic light-by-light scattering contribution to the muon anomaly, with all systematic errors controlled (below we show the omitted non-leading disconnected diagrams are likely negligible). It appears that this contribution can not ``rescue" the Standard Model (or the E821 experiment).

\begin{figure}
    \centering
    \includegraphics[width=0.4\textwidth]{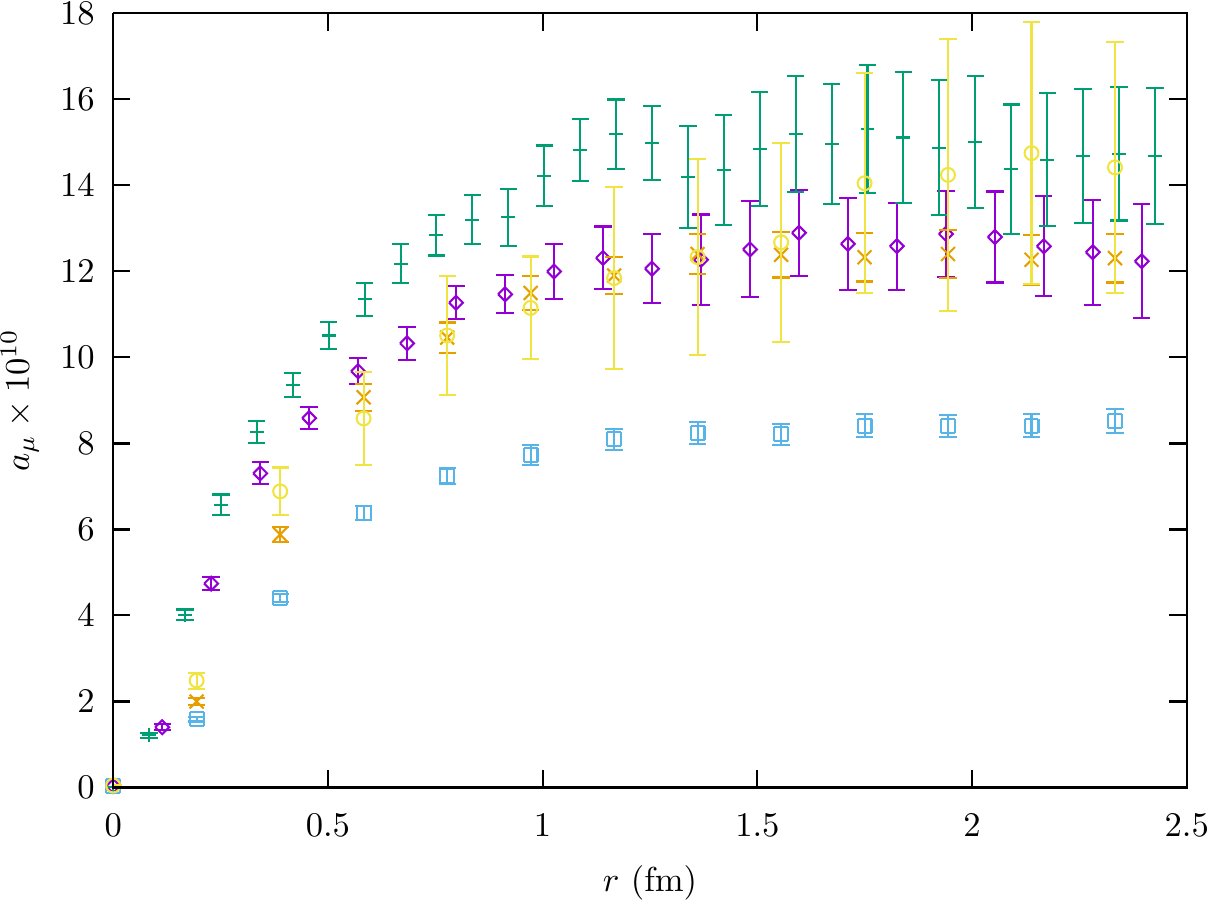}
    \includegraphics[width=0.39\textwidth]{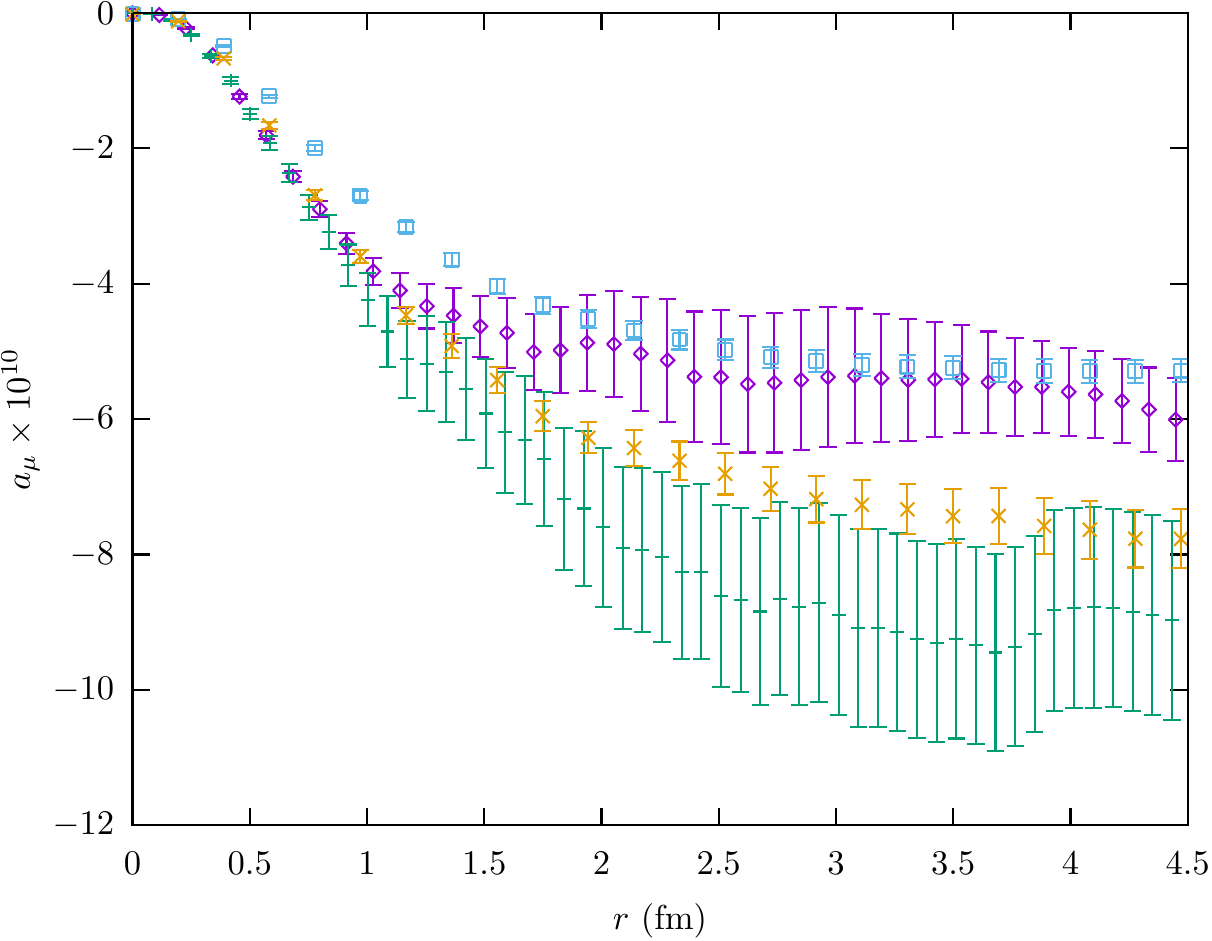}    
    \caption{Cumulative contributions to the muon anomaly, connected (left) and disconnected (right). $r$ is the distance between the two sampled currents in the hadronic loop (the other two currents are summed exactly). $24^3$ IDSDR (squares), $24^3$ IDSDR (squares), $32^3$ IDSDR (crosses), $48^3$ Iwasaki (diamonds), and $64^3$ Iwasaki (plusses).}
    \label{fig:cumulative}
\end{figure}

In fact we can do even a bit better with the data on hand. As seen in Fig.~\ref{fig:cumulative}, which shows the cumulative sum of all contributions up to a given separation of the two sampled currents in the hadronic loop, the total connected contribution saturates at a distance of about 1 fm for all ensembles. This suggests the region $r\simge 1$ fm adds mostly noise and little signal, and the situation gets worse in the limits. A more accurate estimate can be obtained by taking the continuum limit for the sum up to $r=1$ fm, and above that by taking the contribution from the relatively precise $48^3$ ensemble. We include a systematic error on this long distance part since it is not extrapolated to $a=0$. The infinite volume limit is taken as before. This procedure yields $a_\mu^{\rm cHLbL}=27.61(3.12)(0.32)\times 10^{-10}$, with a statistical error that is roughly $2\times$ smaller and small systematic error. Unfortunately a similar procedure for the disconnected diagram is not reliable, as can be seen in the right panel of Fig.~\ref{fig:cumulative}. The curves do not saturate at 1 fm, but instead tend to increase significantly up to 2 fm, or more. Once the cut moves beyond 1 fm it is no longer effective. The different behavior between the two stems from the different sampling strategies used for each~\cite{Blum:2015gfa}. Using the improved connected result, we find our final result for QED$_L$, 
\begin{equation}
    a_\mu^{\rm HLbL} = (7.41\pm6.33)\times 10^{-10},
\end{equation}
where the error is mostly statistical and includes a small systematic, added in quadrature, for the hybrid continuum extrapolation of the connected diagram.

\section{QED$_\infty$ Method}

\begin{figure}
    \centering
    \includegraphics[width=0.4\textwidth]{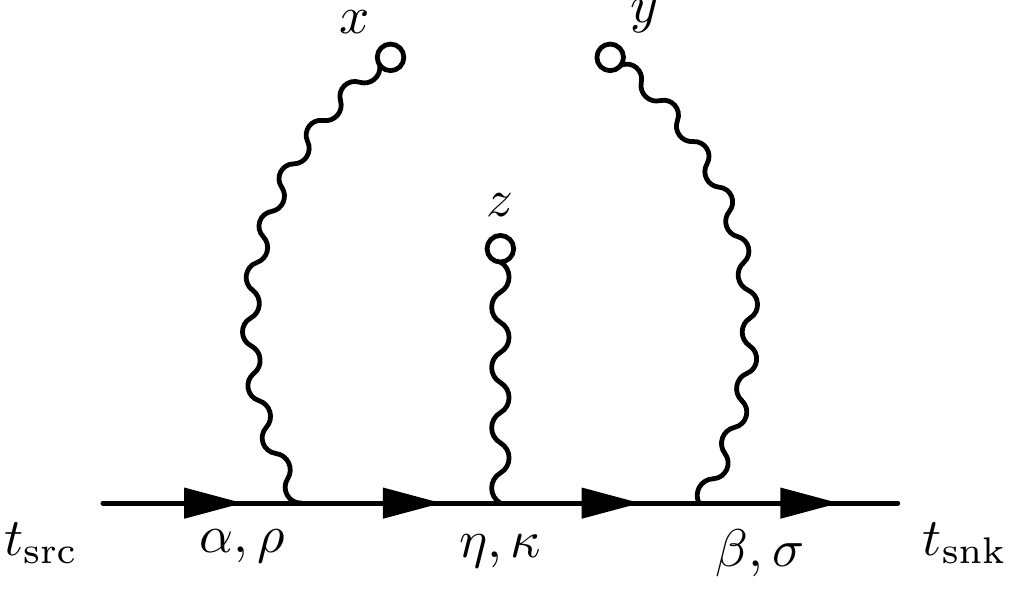}
    \caption{The QED part of the light-by-light scattering amplitude, computed in infinite volume, in the continuum limit. An analytic integral expression as a function of the coordinates $x,y,z$ is pre-computed and tabulated for later use \protect\cite{Blum:2017cer}
    with the hadronic amplitude computed on a discrete, finite lattice.}
    \label{fig:qedinf}
\end{figure}

A method to compute the two-loop QED integrals directly in infinite volume and the continuum was first proposed by the Mainz group~\cite{Green:2015sra,Asmussen:2016lse}. This is similar to what is done in the lattice calculation of the hadronic vacuum polarization contribution to the muon anomaly~\cite{Blum:2002ii}. The advantage is that the leading finite volume error is exponentially suppressed instead of $O(1/L^2)$.  Our group subsequently developed a similar method, adding extra terms to reduce the residual scaling errors induced by the hadronic part~\cite{Blum:2017cer}. The key idea of these methods is to pre-compute the QED part shown in Fig.~\ref{fig:qedinf}, as a function of the coordinates $x,y,z$, which lie on the QCD lattice used for the hadronic part.  However, this function is computed using continuum photon and muon propagators evaluated in an infinite space-time volume. This grid, computed in the continuum, is smoothly interpolated for each set of points used to compute the hadronic part.

A test of the method in QED with the loop living on a discrete lattice reproduces the well known perturbative results for loop masses the same as, and $2\times$, the muon mass, respectively~\cite{Blum:2017cer}. 

\subsection{results}

Before discussing preliminary results for QED$_\infty$, we mention we have found generally that the statistical noise associated with the photons grows with the volume. We therefore expect the QED$_\infty$ method to be noisier than QED$_L$, and this is, in fact, the case. In order to combat the problem we introduce another hybrid approach for the long distance contributions. That is, we compute the dominant $\pi^0$ contribution separately and combine with the full lattice value below some cut. This long-distance $\pi^0$ part is calculated from a model (LMD)~\cite{Knecht:1999gb} for now, but eventually will come from a completely separate, and model independent, lattice calculation. Since the model value is in accord with model independent dispersive results, the results shown below are not expected to change when all lattice computations are used.

\begin{figure}
    \centering
    \includegraphics[angle=-0,width=0.42\textwidth]{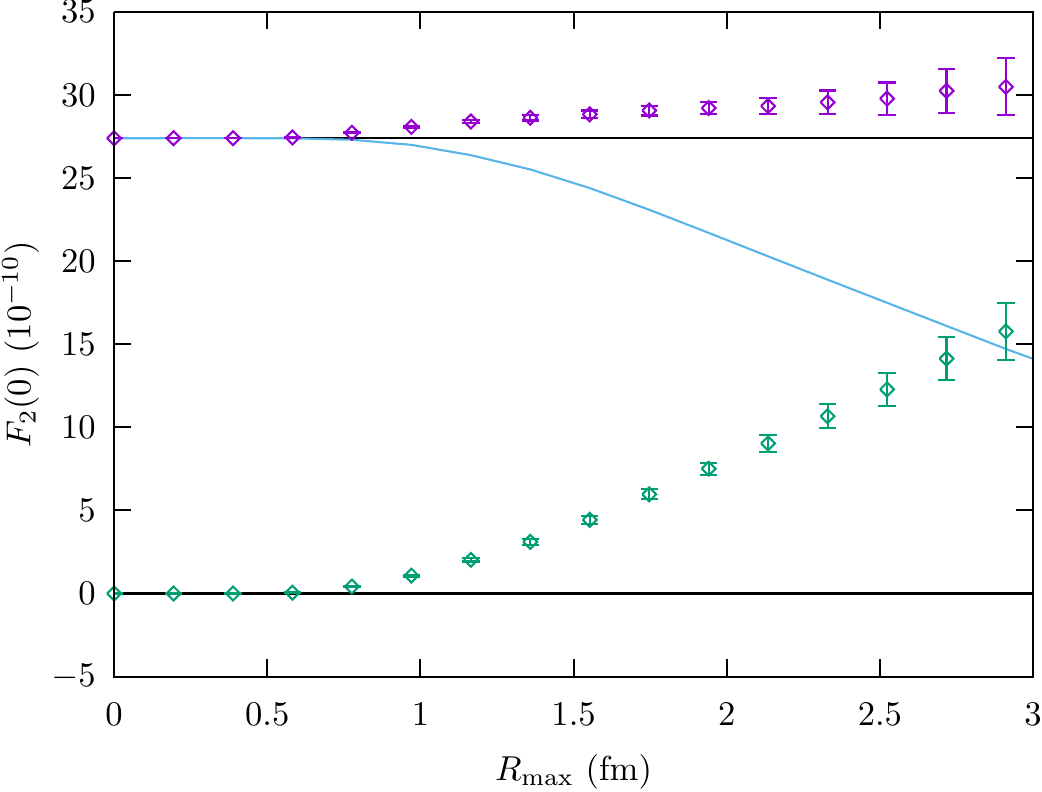}
    \includegraphics[angle=-0,width=0.45\textwidth]{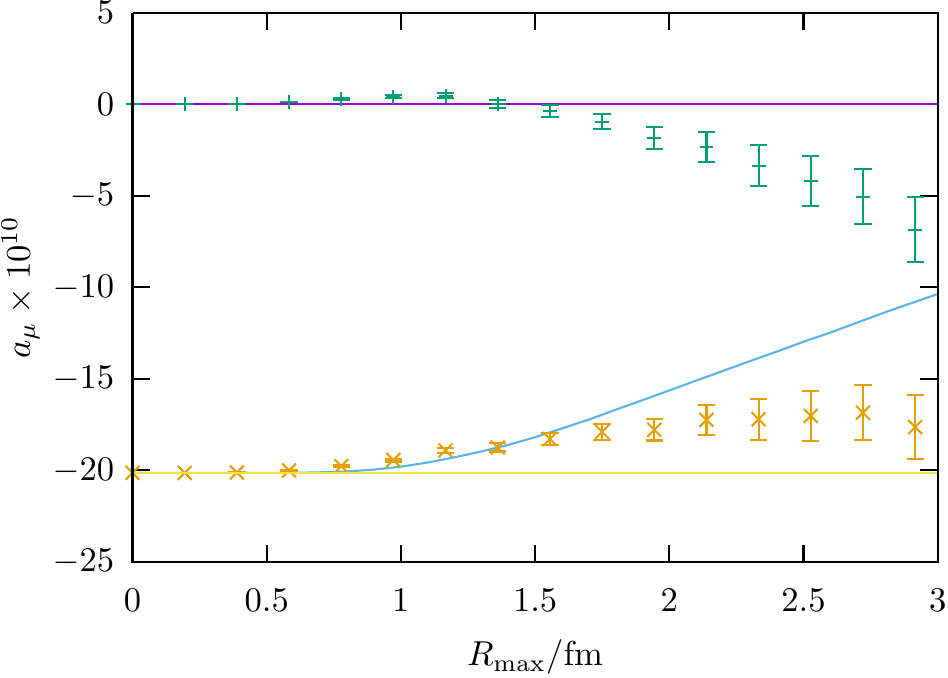}
    \caption{QED$_\infty$ hadronic light-by-light scattering contribution to the muon anomaly ($32^3$, $a^{-1}=1$ GeV). Connected (right) and disconnected (left) are shown. The $\pi^0$ contribution (blue line) is taken from the LMD model for distances between the two sampled points greater than $R_{\rm max}$.}
    \label{fig:qedinf results}
\end{figure}

Figure~\ref{fig:qedinf results} shows both connected and disconnected contributions as the cut between lattice and model contributions is varied. The QCD box in this example is large, roughly 6.4 fm on a side, with a spacing of 0.2 fm ($a^{-1}=1$ GeV). At large $R_{\rm max}$ in the figure, the total is lattice dominated with large uncertainty, and as $R_{\rm max}\to0$, the contribution completely comes from the model. Since the model is only correct at long distance where the $\pi^0$ dominates, at some point the combined result may become constant, yielding an accurate and more precise result than the lattice value alone. One sees that over the range 1-3 fm, lattice (green points) and model results change substantially, but the total remains roughly constant. The respective total values are also roughly compatible with QED$_L$ in the infinite volume and continuum limits. This suggests that residual finite volume and discretization errors are much smaller for QED$_\infty$ (compare to the crosses in Fig.~\ref{fig:cumulative}). This is as expected for the finite volume errors, and it turns out the latter is due to the extra terms added to the QED weighting function (two-loop integral) which vanish in the $a\to0$ limit~\cite{Blum:2017cer}. A similar reduction can be easily seen in the case of pure QED~\cite{Blum:2017cer}. 

\begin{figure}
    \centering
    \includegraphics[angle=-0,width=0.4\textwidth]{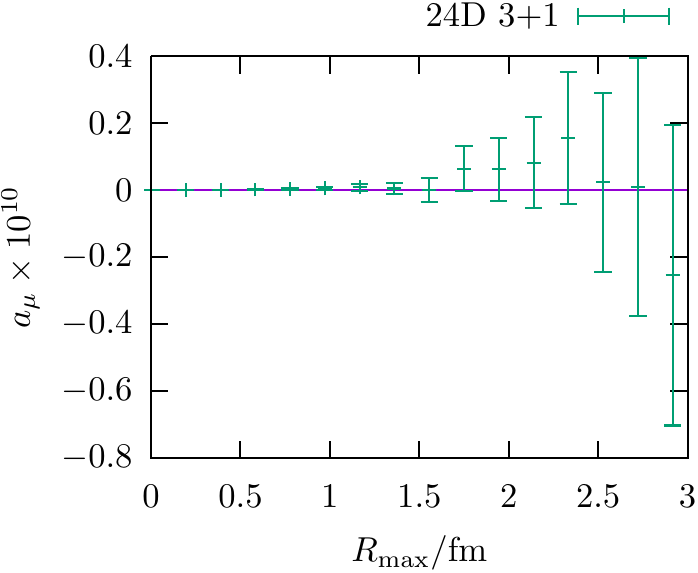}
    \caption{Contribution of non-leading disconnected diagram to the muon anomaly (QED$_\infty$, $24^3$, $a^{-1}=1$ GeV).}
    \label{fig:3+1 disco}
\end{figure}

Finally, we have investigated the size of the next-to-leading disconnected diagram shown in the middle of the top row in Fig.~\ref{fig:disco diags}. As expected and shown in Fig.~\ref{fig:3+1 disco}, this diagram is severely suppressed compared to the leading contributions.

\section{Summary and Outlook}

We have presented preliminary results for the hadronic light-by-light scattering contribution to the muon $g-2$ from Lattice QCD+QED calculations using physical masses, large boxes, and improved measurement algorithms. Both finite volume and infinite volume QED methods are being investigated. For the former, large discretization and finite volume corrections are apparent but under control, and the value in the continuum and infinite volume limits is compatible with previous model and dispersive treatments, albeit with a large statistical error. Despite the large error, which results after a large cancellation between connected and disconnected diagrams, our systematic calculation suggests that light-by-light scattering can not be behind the approximately 3.7 standard deviation discrepancy between the Standard Model and the BNL experiment E821. Future calculations will reduce the error significantly. We have also presented calculations using the QED$_\infty$ method. When combined with a separate calculation of the dominant $\pi^0$ contribution, QED$_\infty$ is statistically effective. It also has much smaller finite volume and discretization errors compared to QED$_L$ for the same QCD box, even for large lattice spacing. These calculations strengthen the exciting test of the Standard Model promised by the new experiments ongoing at Fermilab and planned at J-PARC.
\section*{Acknowledgments}
This work was partially supported by the US Department of Energy. Computations were carried out on the Mira supercomputer at the ALCF at Argonne National Lab.

\end{document}